# Growth of Thin Oxidation-Resistive Crystalline Si Nanostructures on Graphene

Naili Yue,[1] Joshua Myers,[2] Liqin Su,[1] Wentao Wang,[3] Fude Liu,[3] Raphael Tsu,[1] Yan Zhuang,[2] and Yong Zhang[1,*]

[1]Department of Electrical and Computer Engineering

The University of North Carolina at Charlotte, Charlotte, North Carolina 28223, USA

[2]Department of Electrical Engineering

Wright State University, Dayton, OH 45435, USA

[3]Department of Mechanical Engineering

The University of Hong Kong, Hong Kong, China

Abstract

We report the growth of Si nanostructures, either as thin films or nanoparticles, on graphene substrates. The Si nanostructures are shown to be single crystalline, air stable and oxidation resistive, as indicated by the observation of a single crystalline Si Raman mode at around 520 cm$^{-1}$, a STM image of an ordered surface structure under ambient condition, and a Schottky junction with graphite. Ultra-thin silicon regions exhibit silicene-like behavior, including a Raman mode at around 550 cm$^{-1}$, a triangular lattice structure in STM that has distinctly different lattice spacing from that of either graphene or thicker Si, and metallic conductivity of up to 500 times higher than that of graphite. This work suggests a bottom-up approach to forming a Si nanostructure array on a large scale patterned graphene substrate for fabricating nanoscale Si electronic devices.

* E-mail: Yong.Zhang@uncc.edu

Since it was pointed out in 2007 that silicene should have the key electronic properties similar to those of graphene,[1] a great deal of interest has been generated in the growth of silicene. Metallic substrates, mostly Ag but also $ZrB_2$, were typically used for silicene growth. However, the silicene structures grown on these substrates were found to be highly distorted from the ideal (theoretically predicted) low-buckled bilayer structure,[2, 3] and unstable in air.[4] The severe structural distortion renders that the silicene Raman frequency is shifted drastically from theoretically predicted 562 $cm^{-1}$ [5] or 575 $cm^{-1}$ [6] of the ideal structure to around 520 $cm^{-1}$, which is nearly the same as the bulk Si.[4, 7-9] This situation is in stark contrast with that in graphene related structure: no matter how much structure distortion exhibits in graphene the $sp^2$ bonding related G peak at ~1600 $cm^{-1}$ always persists despite the appearance of the $sp^3$ bonding related D peak at ~1300 $cm^{-1}$,[10] whereas in all the reported cases of silicene no Raman mode has been found at frequency close to the predicted value. Additionally, the air stability of silicene remains as a critical issue for this new material to be practically useful. A multi-layer silicene (up to 43 monolayers) grown on Ag was found more stable than a monolayer, but still only lasted for up to 24 hours.[11] Here we report the MBE growth of single crystalline (ultra-) thin Si films on graphene. In the ultra-thin region, for the first time, we observe a Raman mode at ~550 $cm^{-1}$, very close to that of the free-standing silicene. More significantly, we find that the obtained Si structures remain intact even 2-3 years after they were grown, indicating that graphene is unique in serving as an anti-oxidation substrate.

The feasibility of growing silicene on graphene is supported by a few theoretical modeling results: DFT calculations show that the inter-layer binding between silicene and graphene is stronger than the interlayer-layer bonding of graphene layers in graphite;[12] molecular dynamics (MD) simulations indicate that a small Si cluster prefers to form a commensurate monolayer Si raft on the graphite surface,[13] and DFT calculations suggest that silicene structure is energetically more favorable than diamond structure for small Si clusters on graphene.[14] Furthermore, despite the well-known bond length disparity between C and Si structure, it has been predicted that a silicene and graphene could form a commensurate, i.e., nearly lattice matched, heterostrcture in √3x√3R30 stacking with respect to graphene, because of the unique relationship in their bond lengths $d_{Si-Si} \approx \sqrt{3}d_{C-C}$.[12, 15] The graphene-like silicon structure can be viewed as a partially collapsed Si (111) monolayer with its bi-layer separation reduced from $d_{Si-Si}/3$ in the 3D structure to about one half of that in the ideal silicene.[16, 17] Graphene as a substrate is least likely to buckle due to its strong in-plane σ-bonding, and thus less likely to distort the silicene structure. However, its π electrons can be used to facilitate a weak bonding with the epitaxial layer,[18] and yet do not yield significant perturbation to the electronic structure of the epilayer.[12, 15] These considerations motivate us to grow silicene on graphene. Graphene has recently been explored as the substrate for epitaxial growth of $MoS_2$ and $MoSe_2$[19, 20] or a universal buffer layer to grow other semiconductor materials on any substrate without the constrain of lattice matching.[18] Growing Si on graphite or graphene is also of interest to develop low cost Si photovoltaics[21] as well as flexible Si electronics.[22]

As a matter of fact, there has been some past effort to grow Si on graphite before the recent interest in silicene, for instance, fullerene-structured Si nanowires,[23] ultra-thin Si films,[24] thick Si epilayers,[25, 26] and Si nanocrystals.[27] However, such Si materials were often highly defective polycrystalline.[25-27] As expected for polycrystalline Si, the primary Si Raman mode near 520 cm$^{-1}$ was found to be significantly broadened and red shifted.[27] Recently, ultra-thin silicon films deposited on highly oriented pyrolytic graphite (HOPG) and sapphire substrates were reported to exhibit $sp^2$-like bonding in photoemission studies.[28]

In this work, thin Si films were grown on graphite and graphene on $SiO_2$/Si substrates in a MBE system (SVT Associates Inc.) by evaparating bulk Si with an e-beam evaporator. Graphite substrates of a few mm size were cleaved from a large single crystal graphite. Si was deposited in the central region of the small graphite substrate. The typical growth conditions are as follows: growth chamber base pressure being $2x10^{-8}$ Torr; heating the substrate to the growth temperature $T_g$ = 800 or 850 °C and held for 15 minutes; e-beam evaporator running with acceleration voltage 6.07 kV, emission current 150 mA, and filament current 31 A; growth time $t_g$ = 15 or 10 minutes; holding at $T_g$ for 5 minutes; cooling rate 10 °C/min from $T_g$ to 500 °C, then cooled down naturally to room temeorature in the growth chamber.

Surface morphology was charcterized by SEM and AFM. Si particles and thin-film-like structures were found to form on the cleaved graphite surface that exhibited various clean and flat regions more than 10 µm in size. These regions provide high quality single crystalline graphene to serve as template for epitaxial growth. Transferred graphene on other types of susbstrates is likely more defective, either due to the presence of polycrystalline domains or chemical residues associated with the transfer. A few typical SEM images are shown in Figure 1. Figure 1(a) and (b) were taken from sample S1 with $T_g/t_g$ = 800 °C/15 min., showing two areas of different densities of Si particles or islands, roughly 100 – 200 nm in size. Figure 1(c) and (d) from sample S2 with $T_g/t_g$ = 850 °C/10 min., showing one area with very small Si particles in the order of 10 nm, and a thin-film like structure of a few µm in size possibly with embedded small Si particles. The heights of these Si structures were found in the range of 1 to 15 nm measured by AFM, as shown in the two represenative AFM images, Figure 2(a) and (b). Another sample (S3) grown on a graphene/$SiO_2$/Si substrate was examined by TEM, which indicates that Si nanocrystals, typically a few nm in size, were observed on the surface. Figure 2(c) is a low magnification image, showing Si layer deposited on the graphene/SiO$_2$ substrate. Figure 2(d) is a high resolution image with visible Si lattice planes of a single Si nanocrystal, but the graphene layer is too thin to see.

The epitaxial Si structures were characterized by confocal microRaman using a Horiba LabRam HR800 Raman microscope with a 100x lens (NA = 0.9), excited with a 532 nm laser. A sufficiently low laser power (~ 1 mW) was used to minimize heating induced peak shift. Figure 3 shows a few represenative Raman spectra from the Si on graphite samples. Figure 3(a) is from S1 measured on two areas: one with a Si particle and the other a uniform area, compared with a bulk Si. In contrast to the severely distorted Raman spectra reported for Si nanoparticles also grown on graphite,[27] here we have oberved single crystalline Si-like Raman spectra for the epitaxial Si structures, with only a small redshift in the peak position and small broadening in linewidth. Interestingly, the shift of the thin-film area is slightly more than the the particle that is somewhat thicker. Note that despite the expected close lattice matching between graphene and Si (111), the in-plane lattice constant of Si is actually a few percent smaller. It has been documented that 2D films like monolayer MoS$_2$ and WS$_2$ usually form significant chemical bonding with the substrates on which they are grown.[29-31] Given the predicted weak but signifiant chemical bonding between graphene and silicene,[12, 15] we expect that the thin diamond-like Si strcuture could experience some tensile epitaxial strain from the graphite substrate.[12, 15] The strain could qualitatively explain the variation in redshift that is larger for the thinner layer. Also, the expected nonuniform bond lengths along the growth direction, due to the variation of the in-plane lattice constant with the thickness, might also contribute to the small Raman line broadening. In terms of Raman intensity, if we assume that Raman signal is proportional to the sample volume, based on the absorption coefficient of Si ($\alpha \sim 10^4$ cm$^{-1}$ at 532 nm), we can offer a rough estimate for the Si film thickness to be 1.4 nm (i.e., 4-5 monolayers thick using the monolayer thickness of buck Si at 3.15 Å), which is consistent with what we measured with AFM from thin area, such as Figure 2(b).

Based on the phonon frequency change between diamond $F_{2g}$ mode (~1300 cm$^{-1}$) and graphene $E_{2g}$ mode (~1600 cm$^{-1}$), one would expect that the silicene $E_{2g}$ phonon frequency to be roughly in proportion higher than that of bulk Si at ~520 cm$^{-1}$. Indeed, the theoretically predicted value for free-standing silicene is 562 cm$^{-1}$ [5] or 575 cm$^{-1}$.[6] Therefore, the spectra shown in Figure 3(a) are likely of bulk like Si structures. However, at certain locations that appear to have ultra-thin Si films based on the signal strength, we have instead observed a Raman mode at 550.5 cm$^{-1}$, as shown in Figure 3(b) with spectra measured from multiple Si sites and graphite sites. On those Si sites, there is an anti-correlation between the 3D Si peaks near 520 cm$^{-1}$ and the new Si related Raman mode near 550 cm$^{-1}$. We note that these Si related spectra are distinctly different from those of graphite that do not exhibit any well defined feature in the same spectral range. This 550.5 cm$^{-1}$ mode is much closer to the predicted free-standing silicene mode, and the redshift from the theoretical value could be due to the presence of the tensile strain from the substrate.[12, 15]

It is unusual that the Raman spectra of these very thin Si samples remain highly stable after 2-3 years the sample were grown. One would expect that the thin Si structures had been mostly oxidized and converted into $SiO_2$, given the oxidation rate of 11-13 Å in one day[32] or about 2 nm in one month[33]. This is one important indication of the anti-oxidation effect of the graphene substrate. However, more intriging and convincing finding is offered by STM measurements done on one of the samples.

Figure 4 shows the electrical characterization and STM images for three distinctly different regions on sample S1: of no Si growth (i.e., exposed graphite), of ultra-thin Si, and of relatively thick Si. These measurements were acquired using an Agilent AFM 5420 atomic force microscope with a STM nose cone and scanner. The tip was prepared by cutting the wire at a 45º angle prior to lowering into position. The current scans were performed in constant current mode, and STM images were obtained in constant height mode. The I-V curves were taken by bringing the tip into contact with the sample at different selected locations of interest, where the tip was held at a constant position and a voltage sweep was performed while measuring the current. The surface of the graphite substrate away from the growth region was used as one contact, and the tip was grounded. Figure 4(a) is the current map of an area with Si deposition, showing ribbon-like Si structures. The brownish color area is graphite, the lightest color area is the thicker Si, whereas the dark area in between is the ultra-thin, silicene like Si, judged by their I-V characteristics and STM images. Note that the strong current contrast revealed in Figure 4(a) is because of the very large variations in conductivity between the three regions such that, despite attempting to measure in the constant current mode, the system was not able to maintain a constant current. Figure 4(b) contrasts the typical I-V characteristics of the three regions under the contact mode. The graphite region is least conductive, then the thicker Si region, and the silicene-like region is most conductive, with a conductivity of up to 500 times that of the graphite region. For instance, at 3.5 mV the current of the silicene-like region is 370 times that of the graphite region. The high conductivity of the silicene-like region could be due to the charge transfer effect from graphite to silicene.[15] Figure 4(c), (e), and (g) plot the I-V curves of the three regions in an extended voltage range, respectively. Both graphite and silence-like regions show ohmic behavior though with large difference in conductivity, whereas the thicker Si exhibits Schottky junction type characteristic, consistent with literature reports for either graphite or graphene/bulk Si junctions.[34-36] The conductivity change with increasing film thickness is qualitatively consistent with the expectation that beyond two monolayers, the multi-layer silicene or thin-Si film becomes a semiconductor.[37] Figure 4(d), (f), and (h) are the corresponding STM images obtained under ambient condition from the three regions, respectively. They show distinctively different patterns. The pattern of Figure 4(d) resembles that expected for graphite, a triangular lattice,[38] although it is highly distorted, and the bright-spot separation of 2.40 ± 0.43 Å is in good agreement with the lattice constant of graphite at 2.46 Å. Despite the topmost layer of graphite is a graphene layer with a hexagonal structure, the STM image should instead be a triangular lattice, due to the interference of the underneath layer.[38, 39] The patterns in Figure 4(f) and (h) for the Si areas are more regular. They both are triangular lattices, but the spacings are quite different from each other and from that of the graphite region. In Figure 4(f) for the silicene like Si, the pattern is consistent with what is expected for the Si version of graphite,[38] the bright-spot separation are 3.53 Å ± 0.19 Å, somewhat smaller than the silicene lattice constant (about 3.8 Å). The structure revealed by Figure 4(h) for the thicker silicon region shows a bright-spot spacing of 1.93 ± 0.22 Å, which does not match any of the known reconstructed Si surfaces.[40] Nevertheless, it is a total surprise that one could observe the Si (111) by STM in air after the long air exposure of the sample (grown in December, 2011, and measured in August, 2013). The exact underlying structures corresponding to these STM images remain to be confirmed through other means, but the differentiations between them confirm that they exhibit distinctively different material properties.

A freshly cleaved Si (111) surface will undergo surface reconstruction if being kept in high vacuum, otherwise will be oxidized into a $SiO_2$ capping layer. Either case, the surface modification is to remove the dangling bonds or minimize the surface energy. Besides $SiO_2$, hydrogen atoms are often used to passivate the dangling bonds in Si. These processes apply to a thick bulk Si. When the layer is sufficiently thin and electronically coupled to a substrate, charge transfer across the heterostructure interface may drastically change the picture. If a very thin Si slab remains in its idealistic $sp^3$ bonding, it will have one dangling bond on the top layer and one on the bottom layer. There are at least two ways to mitigate the dangling bonds: (1) If the slab is only one monolayer thick, partially collapsing the buckled (111) monolayer will allow the upper and lower dangling bonds to form a partial π bond, yielding the so-called silicene that is in-principle structurally stable, although remains chemically unstable (because the weak partial π bond is susceptible to chemical reaction). In contrast, a fully collapsed diamond (111) monolayer forms a much stronger π bond, namely graphene, thus chemically much more stable. (2) Accepting charge from the substrate to passivate the dangling bonds, which has been shown possible theoretically for a silicene/graphene superlattice.[15] Charge transfer induced passivation has been demonstrated to yield stable inorganic-organic hybrid superlattices with two monolayer thick II-VI slabs in reality.[41] It requires more precise growth control and structure characterization to achieve and confirm the feasibility of growing a single layer silicene. However, a self-passivated ultra-thin Si film or multi-layer silicene could potentially be more useful for practical applications than monolayer silicene, because it retains the basic properties of the Si, most importantly the bandgap,[37] whereas silicene is metallic.

We do not yet understand the exact epitaxial relationship between the epitaxial Si structures and the graphene layer. It might be challenging to grow a large and continuous thin Si film. However, it may not be necessary after all if the goal is to make nanoscale Si devices, because a large film is only needed for the traditional top-down approach. This work suggests the possibility to selectively deposit high quality nanoscale Si structures: silicene, a-few-layer silicene, and Si nanocrystals, using a template of nanoscale graphene structures. One possible way to obtain such a template could be firstly growing an array of SiC nanostructures on a large Si wafer then converting them into graphene nanostructures with a laser beam,[42] followed by the growth of Si nanostructures as demonstrated in this work.

**Acknowledgement** The work at UNCC was supported by ARO/Materials Science (Grant No. W911NF-10-1-0281 and W911NF-18-1-0079, managed by Dr. Chakrapani Varanasi). We thank Dr. Weijie Lu for providing the single crystal graphite, Dr. Chun-Sheng Jiang for helpful discussions on STM, Drs. Kai Wang and Gerd J. Duscher for attempting to identify the epitaxial relationship between the substrate and Si structures. YZ acknowleges the support of Bissell Distinguished Professorship.

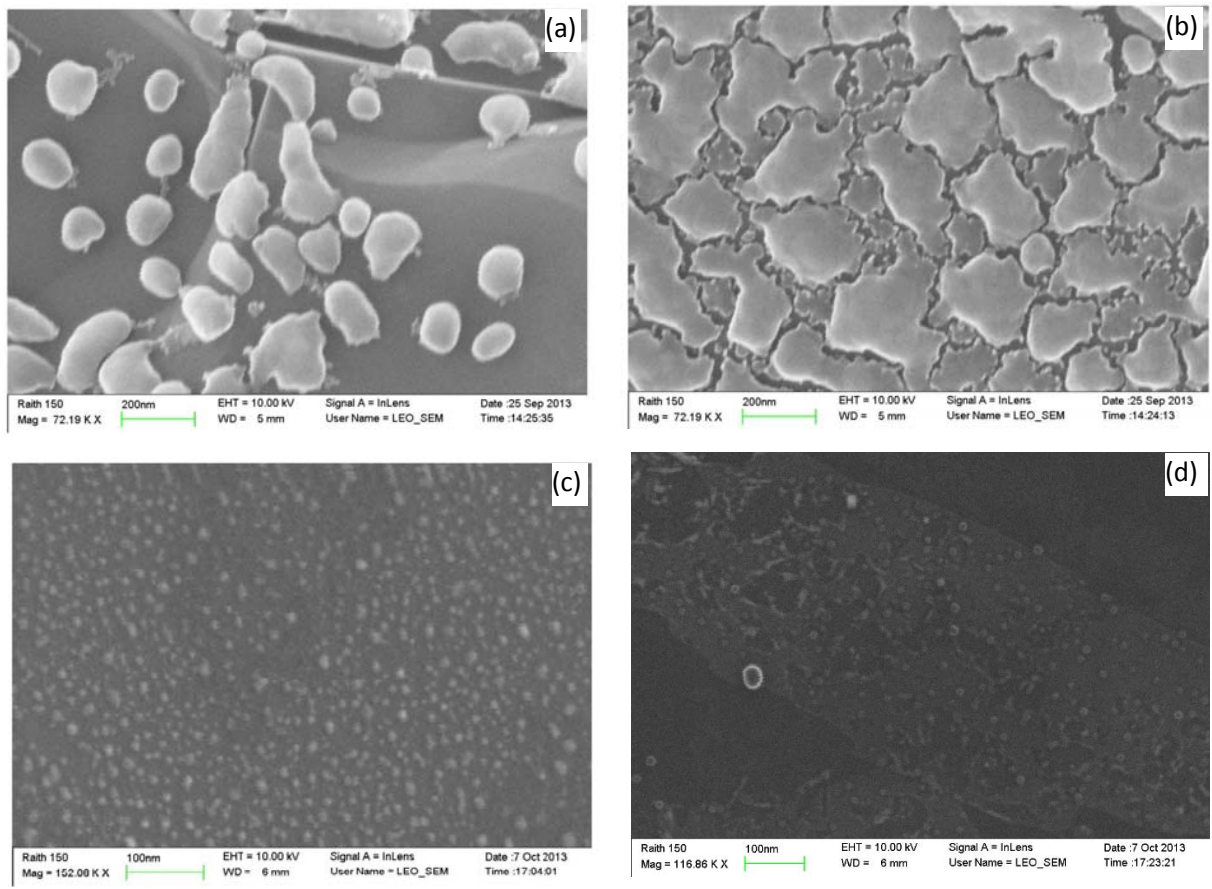

Figure 1. SEM images of epitaxial silicon grown on graphite substrates. (a) and (b) from two areas on S1; (c) and (d) from two areas on S2.

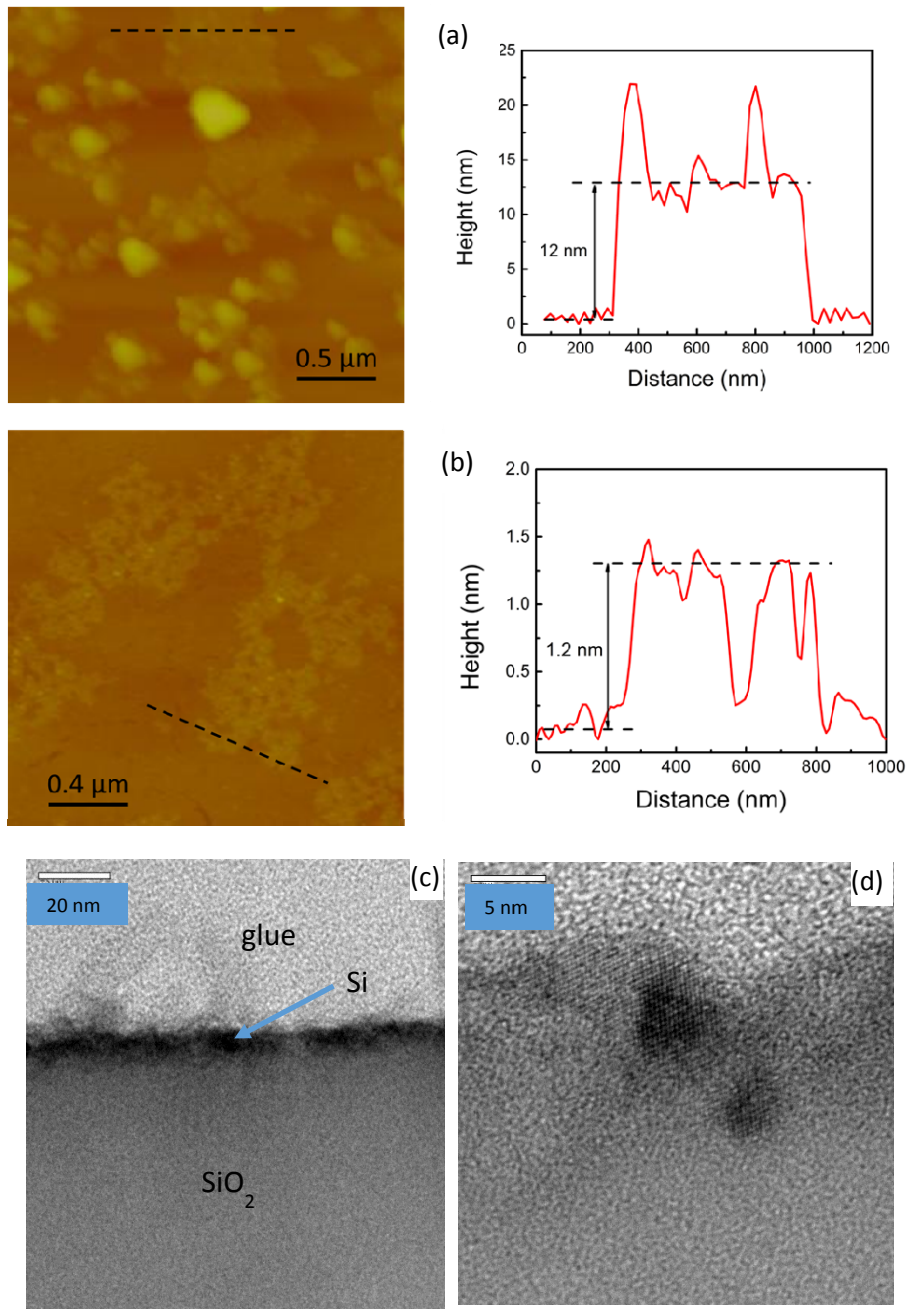

Figure 2. AFM and TEM images of epitaxial thin Si film grown on graphite and graphene. (a) and (b): AFM images from samples S1 and S2, respectively; (c) and (d): TEM images from sample S3.

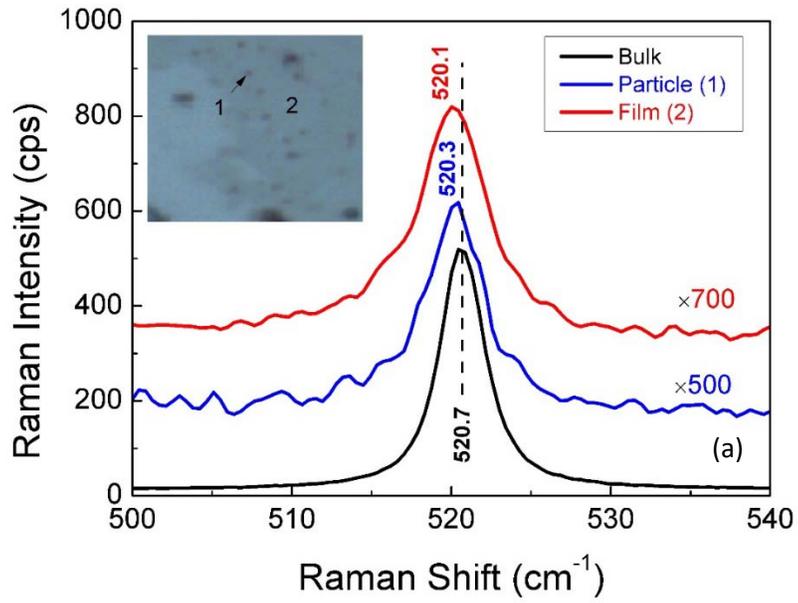
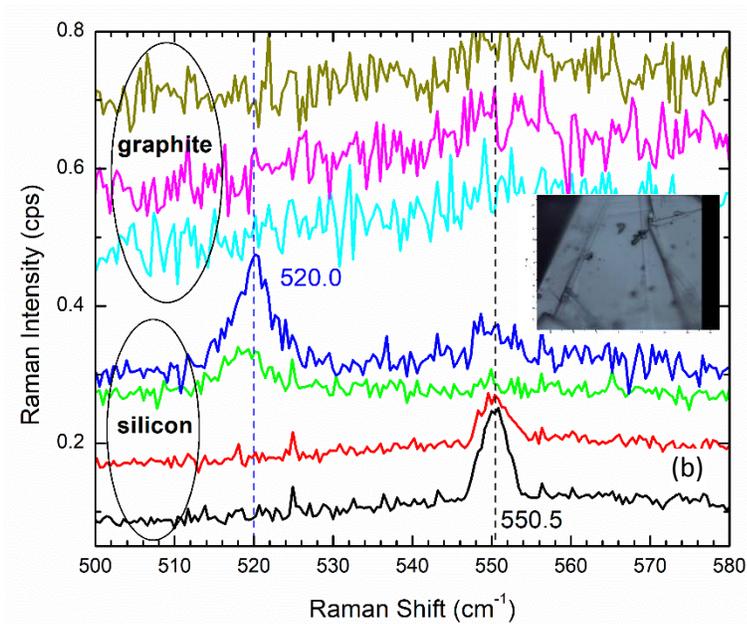

Figure 3. Raman spectra of epitaxial thin silicon on graphene. (a) Spectra from two sites on S1, compared with that of bulk Si; (b) Spectra from multiple sites of thin Si films, compared to graphite spectra (inset: an optical image of the area).

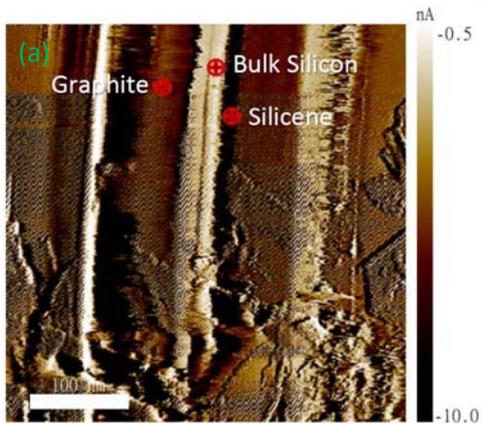
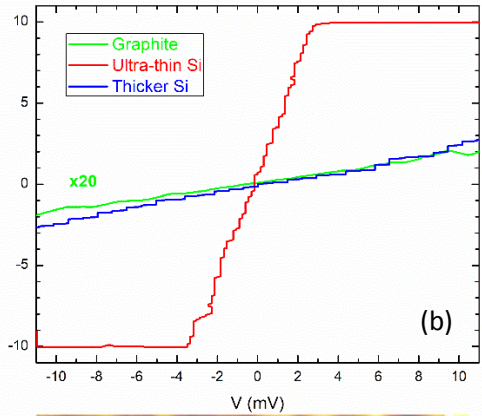
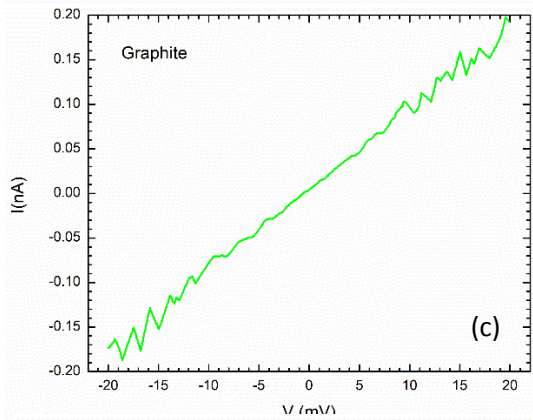
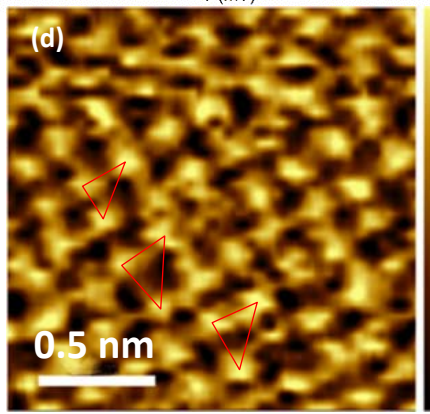
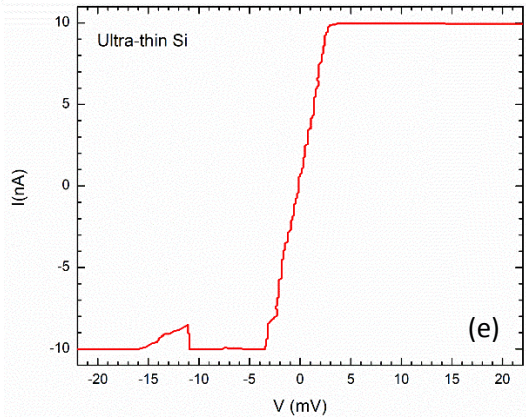
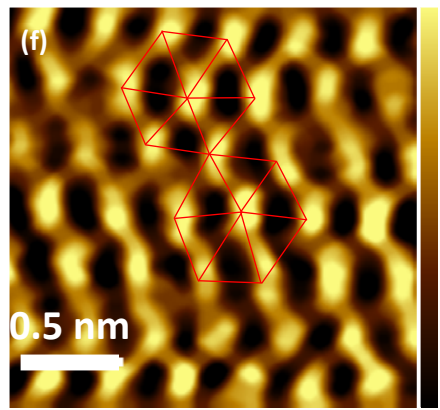
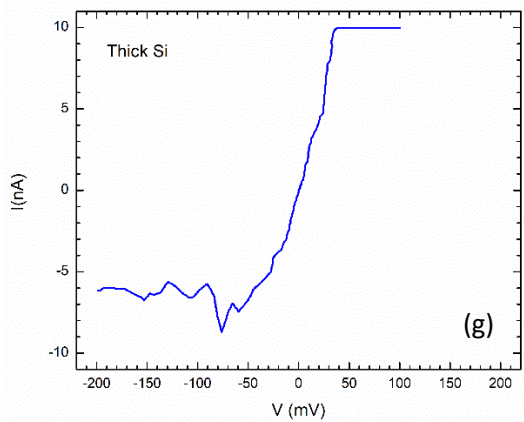
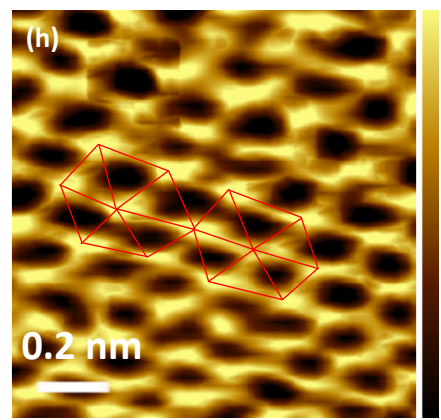

Figure 4. STM images and I-V curves of epitaxial thin silicon on graphite. (a) Current map over a large area containing three types of regions; (b) comparison of I-V curves of the three types of regions under low bias voltages; (c) and (d): I-V curve and STM image of graphite; (e) and (f): I-V curve and STM image of ultra-thin silicon; (g) and (h): I-V curve and STM image of thick silicon.